\documentstyle[aps,multicol,epsfig,eqsecnum]{revtex}
\def\be{\begin{eqnarray}}
\def\ee{\end{eqnarray}}
\def\l{\langle}
\def\r{\rangle}

\begin{document}
\title{ 
Quantum  Disentanglers 
}
\author{
Vladim\'{\i}r Bu\v{z}ek\thanks
{On leave from: 
Institute of Physics, Slovak Academy of Sciences, D\'ubravsk\'a cesta 9,
842 28 Bratislava, Slovakia, and Faculty of Informatics, Masaryk
University, Botanick\'a 68a, 602 00 Brno, Czech Republic. 
}
and Mark Hillery
}
\address{
Department
of Physics and Astronomy, Hunter College, CUNY,
695 Park Avenue, New York, NY 10021, USA\\
}
\date{1 June 2000}

\maketitle
\begin{abstract}
It is not possible to disentangle a qubit in an {\em unknown}
state $|\psi\rangle$ from a set of $N-1$
 ancilla qubits prepared in a specific reference state $|0\rangle$. 
That is, it is not possible to {\em perfectly} perform the  transformation
$\left(|\psi,0\dots,0\r +|0,\psi,\dots,0\r +\dots+
|0,0,\dots\psi\r\right)
\rightarrow |0,\dots,0\rangle\otimes |\psi\rangle$.
The question is then how well we can do?
We  consider a number of different methods of extracting an
unknown state from an entangled state formed from that qubit and
a set of ancilla qubits in an known state.
Measuring the whole system is, as expected, the least
effective method.  We present various quantum ``devices'' 
which disentangle the unknown qubit from the set of ancilla qubits.
In particular, we present the {\em optimal universal} disentangler
which disentangles the unknown qubit with the fidelity which does not
depend on the state of the qubit, and a probabilistic
disentangler which performs the perfect disentangling transformation,
but with a probability less than one.

{\bf PACS number: 03.67.-a, 03.65.Bz}
\end{abstract}
\pacs{03.67.-a  03.65.Bz }

\begin{multicols}{2}


\section{Introduction}
Information encoded in qubits can be used for  reliable quantum communication
or  efficient quantum computing
\cite{Steane1,Gruska}.
This information is
encoded in a quantum state $|\psi (\vartheta,\varphi)\r$ which in the case
of a qubit can be parameterized as 
\be
|\psi (\vartheta,\varphi)\r = \cos\frac{\vartheta}{2} |0\r +
{\rm e}^{i\varphi} \sin\frac{\vartheta}{2} |1\r;
\label{1}
\ee
where $|0\r$ and 
 $|1\r$ are basis vectors of the 2-dimensional space of the qubit
and $0\leq \vartheta \leq \pi$; $0\leq \varphi \leq 2\pi$. 

Qubits are very fragile, that is the state of a qubit can easily be 
changed by the influence of the environment or a random error.
One (very inefficient) way  to protect the quantum information
encoded in a qubit is to measure it. With the help of an optimal
measurement one can estimate the state of a qubit, 
with an average fidelity equal to 2/3 (see below). In this way a quantum
information is transformed into a classical information which can be stored, 
copied, and processed  according the laws of classical physics with
arbitrarily high precision. 
However, in order to utilize the full potential of quantum information
processing we have to keep the information  in  states of
quantum systems, but then we are forced to face the problem of 
decoherence. Recently it has been proposed
that quantum information and quantum information processing can
be stabilized via symmetrization \cite{Barenco}.
In particular, 
the qubit in an unknown state  is entangled with a set of $N-1$ (ancilla)
qubits in
a specific reference state (let us say $|0\r$) so the symmetric state
$|\Psi\r$ of $N$ qubits, 
\be
|\Psi\r\simeq\left(|\psi,0\dots,0\r +|0,\psi,\dots,0\r +\dots+
|0,0,\dots\psi\r\right), 
\label{2}
\ee
is generated.
If we introduce a notation for 
completely symmetric states $|N;l\r$ of $N$ qubits 
with $l$ of them being in the state $|1\r$ and $N-l$ of them in the state
$|0\r$, then the state (\ref{2}) can be expressed in the simple form
\be
|\Psi(\overline{\vartheta},\bar{\varphi})\r=
\cos\frac{\overline{\vartheta}}{2}|N;0\r + {\rm e}^{i\bar{\varphi}}
\sin\frac{\overline{\vartheta}}{2}|N;1\r
\label{3}
\ee
where the parameters $\overline{\vartheta}$ and $\bar{\varphi}$ are specified
by
the relations
\be
\cos\frac{\overline{\vartheta}}{2}
=\frac{\sqrt{N} \cos\frac{{\vartheta}}{2}}
{\sqrt{\sin^2\frac{{\vartheta}}{2}
+ N \cos^2\frac{{\vartheta}}{2}}} \ ; 
\label{4}
\ee
and
$\sin\frac{\overline{\vartheta}}{2}=\sqrt{1-
\cos^2\frac{\overline{\vartheta}}{2}}$, while
$\bar{\varphi}=\varphi$. We see that 
symmetric $N$ qubit state $|\Psi(\overline{\vartheta},\bar{\varphi})\r$
is isomorphic to a single qubit state. But in this case the information is
spread among $N$ entangled 
qubits -  the original quantum information is ``diluted''. Each of the
qubits of the $N$-qubit state (\ref{3}) 
is in the state $\rho_j= \frac{N-1}{N}|0\r\l 0| + 
\frac{(1-\sqrt{N})}{N}(\cos^2
\frac{\overline{\vartheta}}{2}|0\r\l 0| + 
\sin^2\frac{\overline{\vartheta}}{2}|1\r\l 1|) + 
\frac{1}{\sqrt{N}}|\psi
(\overline{\vartheta},\bar{\varphi})\r
\l \psi(\overline{\vartheta},\bar{\varphi})|
$. 

We define the average fidelity between the single  state $\rho_j$ and the
original qubit $|\psi(\vartheta;\varphi)\r$ as
\be
\overline{\cal F}=\int d\,\Omega
\l\psi(\vartheta;\varphi)|\rho_j(\overline{\vartheta},\bar{\varphi}) 
|\psi(\vartheta;\varphi)\r
\label{5}
\ee
where $d \Omega=\sin\vartheta \,d\vartheta d\varphi/4\pi$ 
is the invariant measure on the state space of the original qubit 
(i.e. we assume no {\em prior} knowledge about the pure state
$|\psi(\vartheta;\varphi)\r$).
For this fidelity we find the expression 
\be
\overline{\cal F}_0 = \frac{N^2 -1 - 2 \ln N}{2(N-1)^2}.
\label{5a}
\ee
We see 
that for 
$N=1$ the fidelity $\overline{\cal F}_0$ 
is equal to unity (as it should, because in this
case $|\Psi\r=|\psi\r$) while in the limit $N\rightarrow\infty$ we find
$\overline{\cal F}=1/2$. In fact in this limit density
operators of individual qubits are approximately equal to $|0\r\l 0|$.
In other words, individually the qubits of
the symmetric state $|\Psi(\overline{\vartheta},\bar{\varphi})\r$
in the large $N$ limit do not carry any information about the original
single-qubit state $|\psi\r$. So how can we extract  
the information from the $N$-qubit symmetric state (\ref{3})? The ideal
possibility would be to have have a perfect {\em universal} disentangler
which would perform a unitary transformation
\be
|\Psi(\overline{\vartheta},\bar{\varphi})\r \rightarrow
|\Psi_{ideal}\rangle\equiv
|N-1;0\r\otimes |\psi(\vartheta,\varphi)\r.
\label{7}
\ee
But quantum mechanics does not allow this type of disentangling
transformation \cite{foot1,terno,mor1,mor2}.

While the perfect transformation is impossible, there are a number of
things we can do to concentrate the information from the $N$-qubit state
$|\Psi(\overline{\vartheta},\bar{\varphi})\r$ back into a single qubit. 
In principle, we have the following possibilities: {\bf i)} 
We can either optimally
measure the $N$ qubit state and based on the information obtained prepare a
single-qubit state. {\bf ii)} We can design a quantum disentangler which
would perform a transformation as close as possible to the ideal
disentangling (\ref{7}). In this quantum scenario we have several options
- the process of disentanglement can be input-state dependent. This means
that states (\ref{3}) for some values of the parameters
$\overline{\vartheta}$ and $\overline{\varphi}$ will be disentangled
better than for other values of these parameters. Alternatively,
we can construct a quantum device which disentangles all the state with
the same fidelity. 
{\bf iii)} 
Finally, we propose a probabilistic disentangler,
such that when a specific projective measurement over an ancilla is
performed at the output,  the desired single-qubit state is generated. The
probability of the outcome of the measurement in this case 
is state-dependent.
In what follows we shall investigate all these possibilities.

Before proceeding we note that a different type of disentangler
has been considered by Terno and Mor \cite{terno} - \cite{mor2}.
They considered two different operations.  The first would take 
the state of a
bipartite quantum system and transform it into a state that is
just the product of the reduced density matrixes of the two
subsystems.  The second, which is a generalization of the first,
would again start with a state of a bipartate quantum system,
and map it into a separable state which has the same reduced
density matrixes as the original state.  They showed that
while both of these processes are impossible in general, they
can be realized for particular sets of input states.  An
approximate disentangler of the first type has been considered
by Bandyopadhyay, et.\ al.\ \cite{bandy}. The disentanglers 
we are considering extract, to some degree of approximation,
an unknown state from an entangled state formed from that
state and a known state.

\section{ Measurement scenario} 
Here we first describe a measurement scenario utilizing a set of specific
projection operators. Then we present the optimal measurement-based
approach to quantum disentanglement and we derive an upper bound
on the fidelity of the measurement-based disentangler. 

We utilize  the fact that the $N$ qubit
system prepared in the state 
$|\Psi(\overline{\vartheta},\bar{\varphi})\r$ is isomorphic to a
single qubit. Therefore we first consider a strategy 
based on a 
a projective measurement with two projectors
\cite{Massar,Derka} 
$P_j(\vartheta',\varphi')=|\Xi_j(\vartheta',\varphi')\r
\l\Xi_j(\vartheta',\varphi')|$ ($j=0,1$) with
\be
|\Xi_0(\vartheta',\varphi')\r &=& 
\cos\frac{\vartheta'}{2}|N;0\r + {\rm e}^{i\varphi'}
\sin\frac{\vartheta'}{2}|N;1\r;
\nonumber
\\
|\Xi_1(\vartheta',\varphi')\r &=& 
{\rm e}^{-i\varphi'}\sin\frac{\vartheta'}{2}|N;0\r;
-\cos\frac{\vartheta'}{2}|N;1\r,
\label{8}
\ee      
such that $\l\Xi_j(\vartheta',\varphi')|\Xi_k(\vartheta',\varphi')\r=
\delta_{j,k}$ and $\sum_j P_j(\vartheta',\varphi')=\openone$,
where the angles $\vartheta'$ and $\varphi'$ are chosen randomly if
no {\em prior} information about the measured $N$-qubit 
state is available.

\begin{figure}[t]
\centerline {\epsfig{width=8.0cm,file=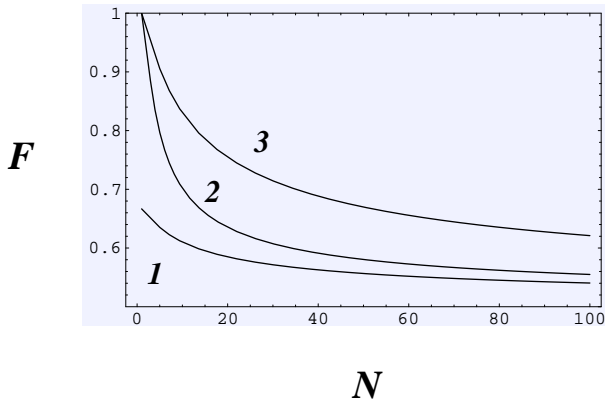}}
\begin{narrowtext}
\bigskip      
\caption{ 
Fidelities of various disentanglers 
 as described in the text.
The line {\bf 1} describes the fidelity $\overline{\cal F}_1$ 
of the measurement-based disentangler 
given
by Eq.~(\ref{12}),  line {\bf 2} is for 
the fidelity  ${\cal F}_2=\gamma_N^2$ 
of the universal optimal disentangler 
given by Eq.~(\ref{15}),
and, finally  line {\bf (3)} is for the mean fidelity of  
the state-dependent disentangler via swapping $\overline{\cal F}_3=
f_N$ 
given by Eq.~(\ref{6}).
 }
\label{fig1}
\end{narrowtext}
\end{figure}

We can use  the result of the
measurement to manufacture a a single-qubit state. Specifically, 
if the result
of the measurement is positive for $P_0$ 
then the single qubit is prepared in the state  
\be
|\eta_0(\vartheta',\varphi')\r = 
\cos\frac{\vartheta'}{2}|0\r + {\rm e}^{i\varphi'}
\sin\frac{\vartheta'}{2}|1\r,
\label{8a}
\ee
 while if the output is positive 
for $P_1$ then the single qubit is prepared in the orthogonal state
$|\eta_1(\vartheta',\varphi')\r$ .  For a particular orientation of
the measurement apparatus (i.e. the angles $\vartheta',\varphi'$)
this measurement-based scenario gives us a single qubit prepared in
the state described by the density operator
\be
\rho^{(meas)}
(\overline{\vartheta},\bar{\varphi};\vartheta',\varphi')=
\sum_{j=0}^1 \left|\l
\Psi
|\Xi_j\r\right|^2
\cdot |\eta_j\r \l\eta_j|  
\label{9}
\ee
After we average over all possible orientations of the measurement
apparatus we obtain on average a single qubit prepared in the state
\be
\rho^{(est)}
(\overline{\vartheta},\bar{\varphi})=
\frac{1}{3}|\psi(\overline{\vartheta},\bar{\varphi})\r
\l\psi(\overline{\vartheta},\bar{\varphi})| + \frac{1}{3}\openone.
\label{10}
\ee
To find the average fidelity of this measurement-based disentangling
procedure we have to evaluate the mean fidelity $\overline{\cal F}_1$,
that is the overlap between the state (\ref{10}) and the
original input state $|\psi(\vartheta,\varphi)\r$ averaged over all
possible orientations of the input qubit:
\be
\overline{\cal F}_1 = 
\int d\, \Omega \l\psi(\vartheta,\varphi)|\rho^{(est)}
(\overline{\vartheta},\bar{\varphi})|\psi(\vartheta,\varphi)\r.
\label{11}
\ee
Taking into account the relation (\ref{4}) we  perform the integration 
in Eq.(\ref{11}) and we find
\be
\overline{\cal F}_1 =
\frac{1}{3}(1+f_N) 
\label{12}
\ee   
where the function $f_N$ reads
\be
f_N=
\frac{N^2 + 4N^{3/2} - 4 N^{1/2} - 1 + 2 N\ln N }
{2(N-1)(N^{1/2}+1)^2}.
\label{6}
\ee   
 For $N=1$:  $\overline{\cal F}_1=2/3$ which is the optimal
fidelity of estimation of the state of a single qubit. From Fig.~\ref{fig1}
we see that the fidelity (\ref{12}) is a decreasing function of $N$
and in the limit $N\rightarrow \infty$ we find $\overline{\cal F}_1=1/2$,
which is equal to the  fidelity of a {\em random} guess associated
with a binary system such as the two projectors under consideration.
In other
words, when the original qubit is diluted into an infinite qubit 
state of the form (\ref{3}) no relevant information can be gained
from  the measurement. The estimated density operator (\ref{10}) in this
case is simply equal to $\openone/2$, which is understandable, because
as we have shown earlier 
in this limit the $N$-qubit state is approximately in the state $|N,0\r$,
so information about the original is ``almost'' totally lost.

\subsection{Optimal measurement scenario}
We now want to find an upper bound $\overline{\cal F}^{max}$ 
 for the average fidelity which can
be achieved by a wide class of measurement-based disentanglement 
procedures.  
We assume that it is {\em a priori} known that 
our $N$-qubit  is prepared in the symmetric state (\ref{2}) with
unknown parameters $\vartheta$ and $\varphi$ associated with a single-qubit
state (\ref{1}). The integration measure on the state  space
of the single qubit is 
$d\Omega = \frac{1}{4\pi}\sin\vartheta d\vartheta d\varphi $ 
and the corresponding
prior probability density distribution on this state space
is constant.

Our strategy is to 
 measure the input state $|\Psi\r$ along the vector
$|\Xi_0\rangle$ [see Eq. (\ref{8})], where the angles
$\vartheta^{\prime}$ and $\varphi^{\prime}$ are chosen according to
the distribution $q(\vartheta^{\prime}, \varphi^{\prime})$, which 
will be left unspecified for the moment.  If the answer is
positive, we produce the output density matrix
$\rho_{0}(\vartheta^\prime,\varphi^\prime)$, 
and if it is negative we produce 
$\rho_{1}(\vartheta^\prime,\varphi^\prime)$, 
where
\begin{equation}
\rho_{j}(\vartheta^\prime,\varphi^\prime)
=\int d\Omega^{\prime\prime}p_{j}(\vartheta^{\prime\prime},
\varphi^{\prime\prime}|\vartheta^{\prime},\varphi^{\prime})
|\eta(\vartheta^{\prime\prime},\varphi^{\prime\prime})\rangle
\langle\eta(\vartheta^{\prime\prime},\varphi^{\prime\prime})|
\end{equation}
with $j=0,1$ and $|\eta\rangle$ given by Eq. (\ref{8a}).
We shall also leave the conditional probabilities, $p_{j}$
unspecified, as this allows us to consider a wide range of
strategies.  For a fixed $|\Xi_0\rangle$, the probability
of the output being $\rho_{0}(\vartheta^\prime,\varphi^\prime)$ 
is $|\langle\Xi_0|
\Psi\rangle|^{2}$ and the probability of it being
$\rho_{1}(\vartheta^\prime,\varphi^\prime)$ 
is $|\langle\Xi_1|\Psi\rangle|^{2}$.  
Averaging over all vectors, $|\Xi\rangle$ gives us
\be
\rho^{(out)}(\overline{\vartheta},\bar{\varphi})
&=&\int d\Omega^{\prime}[|\langle\Xi_0|
\Psi\rangle|^{2}\rho_{0}(\vartheta^\prime,\varphi^\prime)
\nonumber
\\
&+&
|\langle\Xi_1
|\Psi\rangle|^{2}\rho_{1}(\vartheta^\prime,\varphi^\prime)]
q(\vartheta^{\prime},
\varphi^{\prime}) .
\ee
In order to find the average fidelity of the output produced by
this procedure, we compute the fidelity
for a particular input state and average over the input
ensemble
\begin{equation}
\overline{\cal F}=\int d \Omega \langle\psi
(\vartheta,\varphi )|\rho^{(out)}(\overline{\vartheta},\bar{\varphi})
|\psi (\vartheta,\varphi )
\rangle ,
\end{equation}
where $\overline{\vartheta}$ is a function of $\vartheta$
[see Eq.(\ref{4})].
This can be expressed as 
\begin{equation}
\overline{\cal F}=\int d\Omega^{\prime}\int d\Omega^{\prime\prime}
\sum_{j=0}^{1}P_{j}(\vartheta^{\prime\prime},\varphi^{\prime
\prime};\vartheta^{\prime},\varphi^{\prime})f_{j}
(\vartheta^{\prime\prime},\varphi^{\prime\prime};
\vartheta^{\prime},\varphi^{\prime}) ,
\end{equation}
where
\begin{equation}
P_{j}(\vartheta^{\prime\prime},\varphi^{\prime\prime};
\vartheta^{\prime},\varphi^{\prime}) = p_{j}
(\vartheta^{\prime\prime},\varphi^{\prime
\prime}|\vartheta^{\prime},\varphi^{\prime})q(\vartheta^{\prime},
\varphi^{\prime}) ,
\end{equation}
is a normalized joint probability distribution, and
\begin{eqnarray}
f_{0} & = & \int d\Omega|
\langle\Psi |\Xi_0\rangle|^{2}|\langle\psi |\eta\rangle|^{2}
\nonumber \\
f_{1} & = & \int d \Omega|\langle\Psi |\Xi_1\rangle|^{2}|\langle\psi 
|\eta\rangle|^{2} .
\end{eqnarray}
We first note that
\begin{equation}
\int d\Omega^{\prime\prime}p_{j}(\vartheta^{\prime\prime},
\varphi^{\prime\prime};\vartheta^{\prime},\varphi^{\prime})f_{j}
(\vartheta^{\prime\prime},\varphi^{\prime\prime};
\vartheta^{\prime},\varphi^{\prime})\leq
h_{j}(\vartheta^{\prime},\varphi^{\prime}) ,
\end{equation}
where 
\begin{equation}
h_{j}(\vartheta^{\prime},\varphi^{\prime})=\sup f_{j}
(\vartheta^{\prime\prime},\varphi^{\prime\prime};
\vartheta^{\prime},\varphi^{\prime}) ,
\end{equation}
and the supremum is taken over the variables
$\vartheta^{\prime\prime},\varphi^{\prime\prime}$.  We then
have that 
\begin{equation}
\overline{\cal F}\leq \overline{\cal F}^{max}=\sup
[h_{0}(\vartheta^{\prime},\varphi^{\prime})+h_{1}
(\vartheta^{\prime},\varphi^{\prime})] ,
\end{equation}
where the supremum is now taken over $0\leq \vartheta^{\prime}
\leq \pi$ and $0\leq \varphi^{\prime}< 2\pi$. 

In order to calculate this upper bound we must find explicit
expressions for $f_{0}$ and $f_{1}$.  After performing 
the necessary
calculations we find for $\overline{\cal F}^{max}$
the expression
\be
\overline{\cal F}^{max}=
\frac{1}{2}\left[1 + \frac{\sqrt{N}}{(N-1)^3}(N^2-1 - 2N \ln N)\right].
\ee
This fidelity for $N=1$ is equal to 2/3 while in the limit
$N\rightarrow\infty$ is equal to 1/2. For any other $N$ is 
larger than the fidelity $\overline{\cal F}_1$ of the measurement
given by Eq.(\ref{12}) as discussed in our previous example.
Nevertheless, as we will show later it is alway smaller than the
fidelity of the universal quantum device.

\section{Quantum scenario}
In what follows we show 
that a quantum disentangler which preserves quantum
coherences can distill the information
back to a single qubit more efficiently
than can the measurement-based method. As we have already said in the
introduction quantum mechanics does not allow one to construct a 
 perfect disentangler 
which would perform transformation (\ref{7}) for an arbitrary (unknown)
state $|\psi(\vartheta,\varphi)\r$ diluted in the $N$ qubit
symmetric state (\ref{3}). Nevertheless, we can try to design  
optimal disentanglers which perform best under given constraints.

\subsection{State-independent devices }
So let us
assume our quantum disentangler, $D$, is a quantum system with
a $K$-dimensional Hilbert space spanned by basis vectors $|d_k\r$
($k=1,\dots,K$). The disentangler is always initially prepared in   
the state $|d_0\r$, and then it interacts with the $N$-qubit system
in the state (\ref{3}). At the output we want to disentangle the
$N-1$ ancilla qubits from the original qubit, so we expect to
have
\be
|\Psi(\overline{\vartheta},\bar{\varphi})\r |d_0\r
\rightarrow |N-1;0\r \otimes
\sum_{k=1}^K \sum_{j=0}^1 
c_j(\overline{\vartheta},\bar{\varphi}) |j\r |d_k\r.
\label{13}
\ee
As seen from Eq.(\ref{13}) during the disentanglement process the
entanglement between the $N-1$ ancilla qubits and the original qubit
is transferred (swapped) into the entanglement between the original
qubit and the disentangler itself. By tracing over the disentangler
we then expect to obtain the best possible disentangled qubit
in the state $\rho^{(out)}(\overline{\vartheta},\bar{\varphi})$. 
Now we impose several constraints which would
specify what we mean by the optimal covariant (universal) disentangler:
\newline  
{\bf (1)}
The fidelity between the output of the disentangler and the
original state $|\psi(\vartheta,\varphi)\r$ has to be invariant with 
respect to rotations of the original qubit, so the fidelity has to be
input-state independent. This universality of the disentangler would then
guarantee that the information from the symmetric state (\ref{3}) is
extracted for all states equally well.
\newline
{\bf (2)}
We are looking for the {\em optimal} disentangler which would disentangle
the information with the highest fidelity.

Imposing these two conditions we have found the unitary transformation 
which realizes the {\em optimal covariant} disentangler, i.e. which 
disentangle the qubit-state $|\psi\r$ from the $N$-qubit state
$|\Psi\r$ in the optimal and the 
$|\psi\r$-state independent way (see Appendix). 
  This disentangler
is described by the transformation:
\be
|N;0\r |d_0\r &\rightarrow & |N-1;0\r \otimes
\left[ \gamma_N |0\r|d_1\r +\delta_N|1\r|d_2\r\right];
\nonumber
\\
|N;1\r |d_0\r &\rightarrow & |N-1;0\r \otimes
\left[ \delta_N |0\r|d_3\r +\gamma_N|1\r|d_1\r\right];     
\label{14}
\ee
where $|d_j\r$ are three orthonormal basis vectors of the disentangler.
The amplitudes $\gamma_N$ and $\delta_N$ given by the relation
\be
\gamma_N=\left(\frac{N+1}{2(N+1-\sqrt{N})}\right)^{1/2};
\qquad
\delta_N=\sqrt{1-\gamma_N^2}.
\label{15}
\ee

We can directly verify, that the fidelity ${\cal F}_2=
\l \psi(\vartheta,\varphi)
|\rho^{(out)}_d(\overline{\vartheta},\bar{\varphi})|
\psi(\vartheta,\varphi)\r$ is input-state independent and equal to
${\cal F}_2=\gamma_N^2$. Moreover, it can be shown that 
the transformation
(\ref{15}) is optimal, i.e. among all unitary transformations
satisfying the given conditions the transformation (\ref{15})
has the largest fidelity.
We see that
for $N=1$ 
the fidelity ${\cal F}_2=1$, which is obvious, because the original
qubit has not been entangled with ancilla qubits. We plot ${\cal F}_2$ in 
Fig.~\ref{fig1}. We see, that it is {\em always} larger than the fidelity
of the disentanglement via measurement. In the limit
$N\rightarrow\infty$ even the quantum disentangler gives us a totally random
outcome. So in this limit, even optimal quantum entangler on which we
impose the universality condition, is not able to extract information
from the state (\ref{3}).

This is one of the main results of our paper - the optimal covariant quantum
disentangler operates better than if the information is extracted 
(disentangled, distilled) from
the symmetrized state (\ref{3}) with the help the of optimal measurement. 
This is due to the fact that $\overline{\cal F}^{max}\leq {\cal F}_2$.

One can also ask the opposite question, 
how can we generate
out of a qubit in an unknown state $|\psi\r$ the symmetric state 
of the form (\ref{3}). It can be shown that within  quantum mechanics
perfect universal entanglers, which would realize the inverse of the
relation (\ref{7}) do not exist. If one wants to create a state (\ref{3})
from a qubit in an unknown state and $N-1$ ancilla qubits in the known
state $|0\r$ again two scenarios are possible, the measurement-based and
quantum scenarios. It is not surprising that the quantum scenario works
better. We have found the optimal universal (covariant with respect to
rotations of the input qubit) quantum entangler given 
by the transformations:
\be
|0\r |N-1;0\r |e_0\r &\rightarrow & 
\left[ \gamma_N |N;0\r|e_1\r +\delta_N|N;1\r|e_2\r\right];
\nonumber
\\
|1\r |N-1;0\r |e_0\r &\rightarrow & 
\left[ \delta_N |N;0\r|e_3\r +\gamma_N|N;1\r|e_1\r\right];     
\label{16}
\ee
where $|e_k\r$ are three orthonormal basis states of the quantum
entangler, $|e_0\r$ is its initial state and the parameters
$\gamma_n$ and $\delta_N$ are given by Eq.(\ref{15}).
One can check that the fidelity  between the output
of this entangler described by the density operator $\rho^{(out)}_e
(\vartheta,\varphi)$ and the ideally entangled state (\ref{3})
is input-state independent (i.e. does not depend on
the parameters $\vartheta, \varphi$) and is equal to
$\gamma_N^2$. This is the best possible universal (covariant) entangler.

\subsection{State-dependent devices }
The universal disentangler gives a higher fidelity than does
the best measurement-based procedure, but it is not obvious that
this is the best that one can do.  In the case of quantum
cloning, the universal cloners are the ones which maximize
the average fidelity \cite{buzek,gisin}.  As we shall see, however,
in the case of disentanglers this is no longer the case; there
are state-dependent devices which are better.

Consider the general disentangler transformation
\be
|N;0\r |b\r &\rightarrow &|N-1;0\r (|0\r |D_1\r +|1\r |D_2\r )
\nonumber \\
|N;1\r |b\r &\rightarrow &|N-1;0\r (|0\r |D_3\r +|1\r |D_4\r ),
\ee
where the vectors $|b_{jk}\r$, are states of the disentangler itself
and need not be orthogonal.  They must, however, satisfy the 
constraints imposed by the unitarity of the above transformation.
The input state for the device is assumed to be $|\Psi (\overline
{\vartheta},\overline{\varphi})\r$, and the ideal output state,
to which the actual output should be compared, is $|\Psi_{ideal}\r
=|N-1;0\r |\psi (\vartheta ,\varphi )\r$.  The output state is 
calculated by starting with the input state, using the
above transformation, and then tracing over the disentangler to 
obtain an output density matrix, $\rho^{(out)}$. One then finds 
the average fidelity for this process, which we shall call
$\overline{\cal F}_{3}$, from
\begin{equation}
\overline{\cal F}_{3} =\int d\Omega \langle\Psi_{ideal}|
\rho^{(out)}|\Psi_{ideal}\r ,
\end{equation}
Note that we are assuming a specific ensemble of input states; the
probability of the one-qubit state $|\psi (\vartheta ,\varphi )
\rangle$ is assumed to be constant on the Bloch sphere.  Our
result for the average fidelity for a state-dependent device
depends on our choice of input ensemble, while for a 
state-independent device the average fidelity is independent of
this ensemble.  

The calculation of the average fidelity is given 
in the Appendix, and will not be given in
detail.  We find that $\|D_2\|^2 = \|D_3\|^2=0$ and $|D_1\r=|D_4\r$.
This implies that the final state is just a product of the state
of the $N$ particles and the entangler state, which means that
the entangler states can be dropped from the problem.  Therefore,
the transformation which maximizes the average fidelity is just
\be
|N;0\r &\rightarrow &|N-1;0\r |0\r \nonumber \\
|N;1\r &\rightarrow &|N-1;0\r |1\r ,
\label{18a}
\ee
and we have that
\be
|\Psi(\overline{\vartheta},\bar{\varphi})\r \otimes |0\r
\rightarrow |N,0\r \otimes 
|\psi(\overline{\vartheta},\bar{\varphi})\r,
\label{18}
\ee
which is a kind of state swapping transformation.
The average fidelity itself is given by
$\overline{\cal F}_3=f_N$, where the coefficient $f_N$ is given 
by Eq.(\ref{6}). This average fidelity is larger than the
fidelity of the optimal {\em universal} disentangler (see Fig.~\ref{fig1}).
In this case, the fact that 
the universality condition forces us to use an additional quantum device,
the disentangler, with which the  qubit at the output becomes partially
entangled,results in a net loss of information. 
As a result the fidelity of the universal (covariant) entangler is
smaller. 

Analogously, we find that quantum state-dependent entanglement can 
also be performed by  a kind of state swapping transformation, i.e.
\be
|\psi({\vartheta},{\varphi})\r \otimes |N;0\r
\rightarrow |0\r \otimes 
|\Psi({\vartheta},{\varphi})\r.
\label{19}
\ee
with  input-state dependent fidelity 
$|\l\Psi({\vartheta},{\varphi})|
\Psi(\overline{\vartheta},\bar{\varphi})\r|^2$. Nevertheless,
when averaged over all values of $\vartheta,\varphi$ we find the mean 
fidelity of this state-dependent entangler to be equal to $f_N$
which on average is larger than the fidelity of the state-independent
entangler.

\section{Probabilistic disentangler}
Let us examine a simple quantum network which takes as an 
input the $N$-qubit state (\ref{3}). The network is composed 
of a sequence of $N-1$ C-NOT gates
${\cal P}_N=\Pi_{k=1}^{N-1} C_{kN}$ where 
$C_{kl}$ is the C-NOT 
with $k$ being the control bit and $l$ being the target bit.
This sequence of the C-NOT gates acts on the two  vectors $|N;0\r$
and $|N;1\r$ as
\be
\label{20}
{\cal P}_{N}|N;0\r&\rightarrow& |N-1;0\r |0\r
\\
{\cal P}_N |N;1\r&\rightarrow& \frac{1}{\sqrt{N}}
\left(\sqrt{N-1}|N-1;1\r +|N-1;0\r\right) |1\r
\nonumber
\ee
from which it follows that the input vector (\ref{3}) is transformed
as
\be
|\Psi(\overline{\vartheta},\bar{\varphi})\r &\rightarrow&
\frac{\sqrt{N}}{\cal N}
( |v_+\r |\psi(\vartheta,\varphi)\r  \nonumber \\
& &+ \sqrt{N-1}\cos\frac{\vartheta}{2}|v_-\r|0\r )
\label{21}
\ee
where ${\cal
N}=\sqrt{N^2\cos^2\frac{\vartheta}{2}+N\sin^2\frac{\vartheta}{2}}$
is the normalization constant. In Eq.(\ref{21}) we have introduced
two orthogonal vectors of $N-1$ qubits $|v_\pm\r$. 
\be
|v_+\r&=& \frac{1}{\sqrt{N}}\left\{ 
\sqrt{N-1}|N-1,1\r+|N-1,0\r\right\}
\nonumber
\\
|v_-\r&=& \frac{1}{\sqrt{N}}\left\{ 
\sqrt{N-1}|N-1,0\r-|N-1,1\r\right\}
\label{22}
\ee
At the
output of the network a projective measurement on the first $N-1$ 
qubits is performed in order to determine whether they are in the
state $|v_{+}\r$ or $|v_{-}\r$. 
If the result $|v_+\r$ is obtained, then the $N$th
qubit is in the desired state $|\psi(\vartheta,\varphi)\r$. 
The probability of this outcome is given by
\be
P_{|v_+\r}=\frac{1}{N\cos^2\frac{\vartheta}{2}
+\sin^2\frac{\vartheta}{2}} .
\label{23}
\ee
This probability is input-state-dependent, and it decreases with
$N$.  

There is a difference between this probabilistic process and those
considered previously, such as probabilistic cloning \cite{duan}. 
Those  only work for set of input states which is finite.  The
process considered above, however, works for a continuous, and hence
infinite, set of input states.  It, in fact, works for all input
states of the type we are considering.  Therefore, we can conclude
that the range of applicability of probabilistic devices depends on
the process being considered. 

\section{Conclusion}
We have considered a number of different methods of extracting an
unknown state from an entangled state formed from that state and
a known state.  Measuring the state is, as expected, the least
effective method.  In the case of quantum devices, the universal
device was not best one, at least if average fidelity is used
as the criterion.  Probabilistic quantum devices were seen to
work very well for this operation in that they can be used for
the entire set of input states.

\acknowledgements
This work was supported by the National Science Foundation
under grant PHY-9970507, by the IST project EQUIP under the contract
IST-1999-11053 and by  the CREST, Research Team
for Interacting Career Electronics.

\section*{Appendix: Proof of optimality}
Let us consider the optimal quantum disentangler 
which acts as close as possible to the ideal transformation
(\ref{7}). The disentangler maps  the space spanned
by the vectors $|N;0\rangle$ and $|N;1\rangle$, 
into the 
space spanned by $|N-1;0\rangle|1\r$ and $|N-1,0\r|1\rangle$.  This
suggests that we consider a transformation of the following
form
\begin{eqnarray}
|N;0\rangle |d_0\r &\rightarrow & |N-1;0\r\left(|0\rangle |D_{1}\rangle +
|1\rangle |D_{2}\rangle \right), \nonumber \\
|N;1\rangle |d_0\r &\rightarrow & |N-1;0\r\left(|0\rangle |D_{3}\rangle +
|1\rangle |D_{4}\rangle\right), 
\label{A.1}
\eqnum{A.1}
\end{eqnarray}
where $|d_0\r$ is the initial state of the disentangler which is supposed
to be the same for all inputs and $|D_j\r$ ($j=1,\dots,4$) are some
unnormalized disentangler state-vectors. Our task is to determine
these vectors.

Unitarity immediately implies that
\begin{eqnarray}
\| D_{1}\|^{2}+\| D_{2}\|^{2}&=&1 
\nonumber
\\
 \| D_{3}\|^{2}
+\| D_{4}\|^{2}&=& 1
\label{A.2}
\eqnum{A.2}
\\
\langle D_{1}|D_{3}\rangle + \langle D_{2}|D_{3}\rangle
&=& 0 .
\nonumber
\end{eqnarray}
We shall now use our disentangler transformations (A.1) 
to calculate 
the fidelity of the actual output to the ideal output (\ref{7})
The input of the disentangler  is given by Eq.~(\ref{3}).
If we introduce a notation 
$\overline{\alpha}= \cos\frac{\overline{\vartheta}}{2}$ and 
$\overline{\beta}= {\rm e}^{i\varphi}\sin\frac{\overline{\vartheta}}{2}$ 
we can write the 
result of the transformation (A.1) 
\begin{eqnarray}
|\Psi_{out}\rangle & = & 
|N-1;0\r \otimes \left[
\overline{\alpha}\left(
|0\rangle |D_{1}\rangle +
|1\rangle |D_{2}\rangle \right)\right.
 \nonumber \\
 && + \left.
\overline{\beta}
\left(|0\rangle |D_{3}\rangle +
|1\rangle |D_{4}\rangle\right) \right].
\label{A.3}
\eqnum{A.3}
\end{eqnarray}
We now use this expression to find the output density
matrix and trace out the disentangler itself.  We
define the $N$-qubit output density matrix to be
\begin{equation}
\rho_{out}={\rm Tr}_{\rm disentangler}(|\Psi_{out}\rangle\langle
\Psi_{out}|) .
\label{A.4}
\eqnum{A.4}
\end{equation}
The output fidelity is given by
\begin{equation}
{\cal F}=\langle\Psi_{ideal}|\rho_{out}|\Psi_{ideal}\rangle ,
\end{equation}
where $|\Psi_{ideal}\r$ is given by Eq.~(\ref{7}). 
If we denote $\alpha=\cos\frac{\vartheta}{2}$ and $\beta={\rm e}^{i\varphi}
\sin\frac{\vartheta}{2}$ we can express this fidelity as
\begin{eqnarray}
{\cal F} & = & \frac{1}{(N|\alpha |^{2}+|\beta|^2)}\{ N|\alpha |^{4}\| 
D_{1}\|^{2}+|\beta |^{4}\| D_{4}\|^{2} 
\nonumber \\
&+&|\alpha |^{2}|\beta |^{2}[\| D_{3}\|^{2} 
 +N\| D_{2}\|^{2}+\sqrt{N}(\langle D_{4}|D_{1}\rangle
+\langle D_{1}|D_{4}\rangle )] 
\nonumber
\\
&+&\alpha^{\ast}\beta 
|\alpha |^{2}(\sqrt{N}\langle D_{1}|D_{3}\rangle
  +N\langle D_{2}|D_{1}\rangle )
\nonumber
\\
&+&\alpha\beta^{\ast}
|\alpha |^{2}(\sqrt{N}\langle D_{3}|D_{1}\rangle + N
\langle D_{1}|D_{2}\rangle )
\nonumber
\\
&+&\alpha^{\ast}\beta |\beta
|^{2}(\sqrt{N}\langle D_{2}|D_{4}\rangle 
  +\langle D_{4}|D_{3}\rangle)
\nonumber 
\\
&+& \alpha\beta^{\ast} 
|\beta |^{2} (\sqrt{N}\langle D_{4}|D_{2}\rangle +
\langle D_{3}|D_{4}\rangle ) \nonumber \\
 & + &(\alpha^{\ast})^{2}\beta^{2}\sqrt{N}\langle D_{2}|
D_{3}\rangle 
+\alpha^{2}(\beta^{\ast})^{2}\sqrt{N}
\langle D_{3}|D_{2}\rangle \}.
\label{A.5}
\eqnum{A.5}
\end{eqnarray}

From this point on we will study two separate cases. Firstly, we will
prove optimality of the universal disentangler and then the
optimality of the state-dependent disentangler.

\subsection*{A.1 Universal disentangler}
Demanding that the fidelity be independent of phases of
$\alpha$ and $\beta$ we find that
\begin{eqnarray}
\sqrt{N}\langle D_{1}|D_{3}\rangle +N\langle D_{2}|
D_{1}\rangle &=& 0 
\nonumber
\\
\langle D_{3}|D_{2}\rangle &=&0
\label{A.6}
\eqnum{A.6}
\\
\sqrt{N}\langle D_{2}|D_{4}\rangle +N\langle D_{4}|
D_{3}\rangle &=& 0   .
\nonumber
\end{eqnarray}
Assuming these conditions to be satisfied the fidelity 
becomes
\begin{eqnarray}
\label{A.7}
\eqnum{A.7}
{\cal F} & = & \frac{1}{(N|\alpha |^{2}+|\beta|^2)}\{ N|\alpha |^{4}\| 
D_{1}\|^{2}+|\beta |^{4}\| D_{4}\|^{2} 
 \\
&+&|\alpha |^{2}
|\beta |^{2}[\| D_{3}\|^{2} 
 +N\| D_{2}\|^{2}
\nonumber
\\
&+&\sqrt{N}(\langle D_{4}|D_{1}\rangle
+\langle D_{1}|D_{4}\rangle )] \} .
\nonumber
\end{eqnarray}
In order for this to be independent of $\alpha$ and
$\beta$, the term in brackets must be proportional to
\begin{equation}
\label{A.8}
\eqnum{A.8}
(N|\alpha |^{2}+|\beta |^{2})
=N|\alpha |^{4}+(N+1)|\alpha |^{2}|\beta |^{2}+|\beta |^{4} .
\end{equation}
Comparing Eqs. (A.7) and (A.8) we find that
\begin{eqnarray}
\| D_{1}\| & = &\|D_{4}\| \nonumber \\
\label{A.9}
\eqnum{A.9}
(N+1)\| D_{4}\|^{2} & = &
\| D_{3}\|^{2}+N\| D_{2}\|^{2}
\\&&
+\sqrt{N}(\langle D_{4}
|D_{1}\rangle +\langle D_{1}|D_{4}\rangle ) .
\nonumber
\end{eqnarray}
Combining these requirements with those imposed by
unitarity we conclude that
\begin{equation}
\| D_{3}\|^{2} =\| D_{2}\|^{2} = 1-\| D_{4}\|^{2} ,
\label{A.10}
\eqnum{A.10}
\end{equation}
and ${\cal F}=\| D_{4}\|^{2}$.  This means that in order to
maximize ${\cal F}$, we must maximize $\| D_{4}\|^{2}$.

Our first step in accomplishing this is to note that
by combining the results of Eqs. (A.9) and
(A.10) we have that
\begin{equation}
(N+1)+2\sqrt{N} x \|D_{4}\|^{2}=2(N+1)\| D_{4}\|^{2} ,
\label{A.11}
\eqnum{A.11}
\end{equation}
where
\begin{equation}
x=\frac{\langle D_{4}|D_{1}\rangle +\langle D_{1}|
D_{4}\rangle }{2\| D_{4}\|^{2}} ,
\label{A.12}
\eqnum{A.12}
\end{equation}
and $-1\leq x\leq 1$.  Solving for $\| D_{4}\|^{2}$
we find that
\begin{equation}
\| D_{4}\|^{2}=\frac{N+1}{2(N-\sqrt{2}x)} ,
\label{A.13}
\eqnum{A.13}
\end{equation}
which, assuming $N\geq 2$, 
 is greatest when $x=1$.  This implies that
$|D_{1}\rangle = |D_{4}\rangle $ and that
\begin{eqnarray}
\| D_{4}\|^{2} &=& \frac{N+1}{2(N+1-\sqrt{N})} \nonumber \\
\| D_{3}\|^{2}&=&\| D_{2}\|^{2}=\frac{N+1-2\sqrt{N}}
{2(N+1-\sqrt{N})} .
\label{A.14}
\eqnum{A.14}
\end{eqnarray}
Imposing now the conditions on inner products we find that
\begin{equation}
\langle D_{3}|D_{4}\rangle = \langle D_{2}|D_{4}\rangle =0.
\label{A.15}
\eqnum{A.15}
\end{equation}

We can summarize our results in the following way.  Let
$\{ d_{j}|j=1,2,3\}$ be a set of three orthonormal vectors
and define two parameters $\gamma_N$ and $\delta_N$ given by
Eq.~(\ref{15})
we then have that
\begin{eqnarray}
|D_{4}\rangle &=& |D_{1}\rangle = \gamma_N |d_{1}\rangle
\nonumber \\
|D_{2}\rangle &=& \delta_N |d_{2}\rangle \nonumber \\
|D_{3}\rangle &=& \delta_N |d_{3}\rangle ,
\label{A.16}
\eqnum{A.16}
\end{eqnarray}
and the universal {\em optimal} 
disentangler transformation is given explicitly by
Eq.~(\ref{14}).

\subsection*{A.2 Input-state dependent disentanglers}
In order to find the optimal input-state dependent
disentangler we find the explicit form of the transformation
(A.1) such that the {\em averaged} fidelity $\overline{\cal F}
=\int d\Omega {\cal F}$ (with ${\cal F}$ given by Eq.~(A.5))
is maximized. Here, as usually, 
 the integration measure is $d\Omega=\sin\vartheta
d\vartheta\ d\varphi/4\pi$. Therefore after the integral over
the phase $\varphi$ is performed we can write the average fidelity
as
\begin{eqnarray}
\label{A.17}
\eqnum{A.17}
\overline{\cal F} &=& \frac{1}{2}\{ \xi_1 N \|D_1\|^2 + \xi_2 \| D_4\|^2
\\
&+& \xi_3[ \|D_3\|^2 + N\|D_2\|^2 +\sqrt{N}(\langle D_1|D_4\rangle
+\langle D_4|D_1\rangle )]\}
\nonumber
\end{eqnarray}
with
\be
\label{A.18}
\xi_1&=&\int_0^\pi\frac{\sin\vartheta d\vartheta}
{N \cos^2\frac{\vartheta}{2}+\sin^2\frac{\vartheta}{2}}
 \cos^4\frac{\vartheta}{2}
\nonumber
\\
\xi_2 &=& \int_0^\pi\frac{\sin\vartheta  d\vartheta}
{N \cos^2\frac{\vartheta}{2}+\sin^2\frac{\vartheta}{2}}
 \sin^4\frac{\vartheta}{2}
\eqnum{A.18}
\\
\xi_3&=&\int_0^\pi\frac{\sin\vartheta  d\vartheta}
{N \cos^2\frac{\vartheta}{2}+\sin^2\frac{\vartheta}{2}}
 \sin^2\frac{\vartheta}{2}
\cos^2\frac{\vartheta}{2}
\nonumber
\ee
After the integration over the parameter $\vartheta$ we find
\be
\label{A.19}
\xi_1&=&\frac{3 - 4 N + N^2 + 2\ln N}{(N-1)^3}
\nonumber
\\
\eqnum{A.19}
\xi_2&=& \frac{-1 + 4N - 3N^2 + 2N^2 \ln N}{(N-1)^3}
\\
\xi_3&=& \frac{-1 + N^2 - 2N \ln N}{(N-1)^3}
\nonumber
\ee

From the unitarity of the disentangling transformation
it follows that $\|D_2\|^2=1-\| D_1\|^2$ and
$\|D_3\|^2=1-\| D_4\|^2$. When we introduce the  notation 
\begin{equation}
u=\frac{\langle D_{4}|D_{1}\rangle +\langle D_{1}|
D_{4}\rangle }{2\|D_1\| \, \| D_{4}\|} ,
\label{A.20}
\eqnum{A.20}
\end{equation}
where $-1\leq u\leq 1$,  and $\eta_1=\|D_1\|^2$; $\eta_4=\|D_4\|^2$
we can rewrite the average fidelity (A.17) as
\be
{\cal F} &=& \frac{1}{2}[\eta_1 N (\xi_1-\xi_3) +  \eta_4(\xi_2-\xi_3)
\nonumber
\\ 
     &+& 2\sqrt{N}\xi_3 u \sqrt{\eta_1\eta_4} + \xi_3(1+N)].
\label{A.21}
\eqnum{A.21}
\ee
Taking into account that $\xi_1>\xi_3$ and $\xi_2>\xi_3$ we 
easily find that the maximum of the mean fidelity (A.21) is achieved
for $u=1$ and $\eta_1=\eta_4=1$. In this case we rewrite (A.21) 
 as
\be
{\cal F}= \frac{1}{2}[\xi_1 N + \xi_2 + 2 \sqrt{N}\xi_3].
\label{A.22}
\eqnum{A.22}
\ee
When  we substitute into Eq.~(A.22) the explicit expression for the parameters
$\xi_j$ given by Eq.~(A.19) we find that the mean fidelity 
is equal to
the function $f_N$ given by Eq.~(\ref{6}). This exactly is equal to
the mean fidelity of the input-state disentanglement performed via the 
state swapping transformation 
described by Eq.~(\ref{18a}). In fact,
from our conditions $\eta_1=\eta_4=1$ it directly
follows that $\| D_2 \|^2=\| D_3\|^2=0$ while $\| D_1\|^2=\| D_4\|^2=1$.
In addition, from  $u=1$ it follows that $|D_1\r=|D_4\r$, so that the
optimal 
state-dependent disentangling transformation is indeed equal to
Eq.~(\ref{18a}), which we wanted to prove.

\end{multicols}


\begin{thebibliography}{10}
\bibitem{foot1}
We also cannot disentangle the state
$|\Psi(\overline{\vartheta},\bar{\varphi})\r
\l\Psi(\overline{\vartheta},\bar{\varphi})|$ in the form
$\Pi_{j=1}^N \rho_j(\overline{\vartheta},\bar{\varphi})$,
which means that we cannot perform measurements on the $N$ qubits.
\end{thebibliography}
\end{document}